\documentclass[conference]{IEEEtran}
\IEEEoverridecommandlockouts
\usepackage{cite}
\usepackage{amsmath,amssymb,amsfonts}
\usepackage{algorithmic}
\usepackage{graphicx}
\usepackage{textcomp}
\usepackage{xcolor}
\usepackage{lipsum}
\usepackage{subfig}

\def\BibTeX{{\rm B\kern-.05em{\sc i\kern-.025em b}\kern-.08em
    T\kern-.1667em\lower.7ex\hbox{E}\kern-.125emX}}

\begin{document}

\title{A Deep Learning Framework for Predicting Digital Asset Price Movement from Trade-by-trade Data \\
}

\author{\IEEEauthorblockN{1\textsuperscript{st} Qi Zhao}
\IEEEauthorblockA{\textit{New York University} \\
\textit{Leonard N. Stern School of Business}\\
\textit{qz796@stern.nyu.edu}\\}
}

\maketitle
\begin{abstract}
This paper presents a deep learning framework based on Long Short-term Memory Network(LSTM) that predicts price movement of cryptocurrencies based on a trailing window. It is the first to use trade-by-trade data to predict short-term price changes in fixed time horizons. By carefully designing features and detailed searching for best hyper-parameters, the model is trained to achieve high performance on nearly a year of trade-by-trade data. The optimal model delivers stable high performance(over 60\% accurancy) on out-of-sample test periods. In a realistic trading simulation setting, the prediction made by the model could be easily monetized. Moreover, this study shows that the LSTM model could extract universal features from trade-by-trade data, as the learned parameters well maintain their high performance on other cryptocurrency instruments that were not included in training data. This study exceeds existing researches in term of the scale and precision of data used, as well as the high prediction accuracy achieved.
\end{abstract}


\section{Introduction}

Since the introduction of Bitcoin in 2008, numerous blockchain-backed cryptocurrencies have been invented and they have attracted investors through their volatile price movement. Similar to the price of financial assets traded on secondary markets, price of these digital assets also behave in a stochastic way but are more volatile in some sense. Unlike stock or other commodities which have regular trading hours and holidays, cryptocurrencies trades on a 24-hour basis and transactions are done very frequently. That results in a dense transaction record, at the same time a set of very continuous time series data. This continuity on the other hand makes trading cryptocurrencies much more challenging: manually monitoring the price movement and keeping up with sudden price changes becomes ineffective and nearly impossible because of its continuity in activities. Having a automated system to interpret scenarios in real-time and predict the direction of price movement will help investors to navigate through the volatile environment and profit from smart trades. This paper presents a deep-learning architecture that utilizes the Long Short-Term Memory network(LSTM) to draw information from the time series data and prediction future price movement. The trained model is capable of making accurate predictions, thereby empowering users to gain instant insight into this volatile market.

\subsection{The Problem and Current Research}
Many have different views on the predictability of asset prices and the degree of market efficiency.\cite{predictability1} It remains an open question that if one could generate "excess return", the additional return relative to the one generated by the market by predicting future price movement. Efficient Market Hypothesis assumes the price behaves as a random walk, and all information is "priced-in".\cite{predictability2} That is, information has been built into the current price, and therefore knowing price information at any given point will not help on predicting the price in the future. However, some studies showed that the asset prices are predictable to some extent.\cite{predictable1} Many researches have been done in predicting future stock prices or predicting future price movement. In general, there are the traditional statistical parametric approach and the emerging machine learning approach.

Due to the success of the LSTM network on processing sequential data, it is natural for one to apply it on financial time series. Some researchers have attempted to predict the price numerically and try to achieve low prediction error, while others have developed classifiers to predict the up/down price movement and try to achieve high accuracy. Similar efforts have done in forecasting the price of digital assets. Substantial works are done in all fields mentioned above, and they will be discussed in detail in a later section. Unlike researches on stock market which competes performance on benchmark data set, researches in the cryptocurrency market have no firm agreement on data sources or baseline performances. In addition to the limitations imposed by inadequate data samples, some other challenges exist in developing a high-performance model for forecasting the prices of digital assets includes making prediction on a data set with highly skewed class distribution, which may produce high but deceptive accuracy. Moreover, the constantly changing price regime raises obstacles in successfully transferring the learned result into an out-of-sample data set and later delivering steady performance in a real trading environment. Last but not least, solely using price data as input have present inadequate information to predict future price as they are affected by numerous factors.  

\subsection{Contributions}
In order to address issues mentioned above, this study builds models on the largest data set so far with 119,712,824 trade-by-trade data points that ranges from 2019/01/01 to 2019/11/30. Trade-by-trade data is a precise and accurate representation of price in real-time. Cryptocurrency exchanges record all posted orders and transactions electronically. While detailed order book changes can hardly be acquired, historical trade-by-trade data could be extracted through exchange-offered API. The training data used in this study are acquired from the Bitcoin-Tether(BTC-USDT) trading pair on Binance. Among all the cryptocurrencies, Bitcoin (BTC) has the longest history and best liquidity.\cite{liquidity-btc} Tether(USDT) is a digital asset designed to match the value of U.S. Dollar. Its exchange rate with U.S. Dollar maintains very close to one, so in many senses it is viewed as a “stable coin” and plays an important role in cryptocurrency trading.\cite{stablecoin-btc} Among digital asset exchanges, Binance has one of the best market depth for most major cryptocurrencies and also a considerable active user base. These ensures that the best representation of digital asset price movement and the robustness of trained models. The learned result could be easily transfer to other cryptocurrency trading pairs since their data structure is essentially the same.

In addition to the scale and precision of data used, this study also achieves high performance on an out-of-sample test periods over three month from 2019/12/01 to 2020/2/28, by applying the parameters learned from a time period prior to this out-sample-data. This indicates that the model is able to predict unseen, future data accurately. The high accuracy over 60\% exceeds current researches and furthermore the same model maintains its high performance on other cryptocurrency trading pairs without additional training. This shows that the LSTM could extract universal feature from trade-by-trade data and consistently make accurate predictions. Results from this study suggests that the asset prices are predictable to some extent, at least for the digital ones. Last but not least, the framework of this study can easily be utilize in real trading settings, as demonstrated by a trading simulation, to empower investors, traders and liquidity providers.

\subsection{Research Flow}
Rather than focusing on predicting prices numerically, this study chooses to work on classifying the directional movement of future prices and aims for high accuracy. This choice is made due to the fact that having a very close price prediction does not always correlates to profitability. 
\footnote{For example, if a trading strategy is to buy the asset when predicted price is higher and sell the asset when predicted price is lower, then for a current price at 99, a predicted price of 99.2 and an actual price of 98.9 would mean buying and losing money. If the predicted price is 98.1, one would make money while having a much larger prediction error of 0.9 instead of 0.3.} 
However, if one could successfully predict the direction of price movement, then he or she could definitely profit from applying such simple up/down strategy. That is, long the asset if predicted direction is up and short otherwise. In contrast to some previous work which make prediction on a "event-based horizon", essentially to predict the price change in the next few "events", i.e. order book changes or trades, this study emphasizes on making prediction in a fixed time horizon. That is, given input prior to time $t$, the model will predict the price movement in the next few minutes or seconds relative to time $t$. The benefits of making prediction in a fixed time horizon is considerable. First, utilization of such prediction allows more rooms for latency, as compared to event-based predictions, which will require the local system to be on the same state as the exchange server to ensure the least slippage. This is beneficial in trading as will show in a later section. Another benefit of predicting in a fixed time horizon is that the model is not subject to the intensity of market movement. For example, during intensive trading times, more than a thousand trades could take place in one minute, whereas in times with minimal activities, only less than one hundred trades happen per minutes. Therefore, predicting the price after a fixed number of trades or order book changes does not deliver a steady monitoring window because trades do not occur regularly in time. Since no one knows the number of trades in the future, such prediction could hardly be utilized by human traders. By deliver accurate prediction on fixed time horizons, the model presented in this study provides an intuitive understanding of the market to users beyond a signal that could only captured by a machine. In the course of research, the author also observed the optimal hyper-parameters, like the prediction horizon and the look back periods, in achieving the highest prediction accuracy. 

$\textit{Outline: }$ The rest of the paper is organized as follows. Section 2 will discuss relevant literature and present their workflows in detail. Section 3 will discuss the data and the preprocessing measures used in this study specifically. Section 4 will present the model and the process of selecting optimal hyper-parameter. Section 5 will contain the overall results on test periods, a trading simulation, as well as the experiment on other cryptocurrency trading pairs.

\section{Related Work}

The Long Short-Term Memory (LSTM)\cite{LSTM} is an improvement on recurrent neural networks and it has solved the vanishing gradients problem. LSTM has been successfully applied to areas like language modeling\cite{languagemodel}, speech recognition\cite{speechrecog}, and many others that deal with sequential data. In recent years, many researchers have focused their work on using the LSTM to analyze financial data. Many more applied it on processing stock price time series. Among them, \cite{deeplob},\cite{universial} uses LSTM to analyze limit order book(LOB) data and achieved high prediction accuracy on benchmark data set F1-2010. Some others \cite{autoencoders} apply the LSTM combined with auto-encoders to analyze stock price series in order to predict future prices. Besides predicting the price itself, there are also studies\cite{beating}\cite{krauss}\cite{CHIANG} building neural networks to predict the future direction of stock prices and aims to achieve high classification accuracy. 

Compared to substantial works done in predicting stock prices, studies are less in number and data precision when it comes to use machine learning methods to forecast the price of digital assets. However, some researchers are still able to make promising results. There has not been many researches on predicting the direction of price movement of digital assets. \cite{nally-btc} uses the LSTM to predict Bitcoin price movement from daily price data and achieved 52.78\% accuracy. \cite{nips-btc} reported high accuracy over 65 percent on out-of-sample test periods. They used several months of bid-ask price of bitcoin sampled in five second intervals, ranging from 2014 to 2016, and focused their prediction on the predicting the latest price, essentially in the next five seconds. Although their achievement is remarkable, the prediction horizon is too short to have price movement large enough to make a profit. In fact, the market prior to 2017 is far less liquid that it is today, and predicting a less liquid market is easier than to predict liquid ones. Therefore, the high accuracy could potentially be the result of a skewed class distribution, as in a less liquid market, most price change tend to be zero because there is no trade. Nevertheless, their work made remarkable contribution to show that the price of Bitcoin is to some extent predictable.

There are more works on predicting the numerical price of digital assets, mainly Bitcoin. \cite{framework-btc} performed feature selection and applied LSTM on daily price data to predict Bitcoin price. \cite{gru-btc} found that gated recurring unit(GRU) model outperforms other models in predicting bitcoin prices from daily price data. \cite{pricealert-btc} applied LSTM on daily and minute price data to develop an anomaly detection system.

Machine learning methods also have been used on other data sources to forecast Bitcoin prices. \cite{volforecast-btc} forecasts the volatility of Bitcoin using LSTM. \cite{BNN} uses Bayesian neural networks to analyze blockchain data and tries to predict future numerical prices.

Although a lot of works reported remarkable results, unfortunately, there is no benchmark data set or consensus on research method or metrics. Therefore, it is not meaningful to compare accuracy with other works when predicted horizon and data precision are completely different. To the best of the author's knowledge, there is no work that uses trade-by-trade data in larger scale to predict short-term price movements of cryptocurrencies on fixed time horizons. This study is the first to extensively explore the process of modeling trade-by-trade data using an LSTM network in order to achieve predictability in price movements. In particular, this study adopts new practice to create training and validation separation and uses new method to reduce the number of redundant input examples while keeping high coverage on the data. Both is essential in evaluating the performance achieved in training data and making sure it is transferable to an out-of-sample test set.

\section{Data, Feature Engineering}

\subsection{Trade-by-trade Data Overview}

\begin{table}
\caption{A glimpse at trade-by-trade data}
\begin{tabular}{|l|l|l|l|c|}
\hline
\textbf{TradeID} & \textbf{Timestamp} & \textbf{Price} & \textbf{Amount} & \textbf{IsBuyerMaker} \\
\hline 
203767769 & 1578200400437 & 7457.18 & 0.042720 & False  \\
\hline 
203767770 & 1578200400614 & 7457.14 & 0.017739 & True  \\
\hline 
203767771 & 1578200401809 & 7457.17 & 0.107827 & False  \\
\hline 
203767772 & 1578200401811 & 7457.16 & 0.061911 & True  \\
\hline 
203767773 & 1578200402497 & 7457.22 & 0.008068 & False  \\
\hline 
\end{tabular}
\vspace{-10pt}
\end{table} 

The trade-by-trade data for training, presented in TABEL I, spans all trades on BTC-USDT trading pair from 2019/01/01 to 2019/11/30, UTC time. There are 119,712,824 trades in total. Each data point in the trade-by-trade data represents one trade that has occurred on the exchange. The details of each trade is represented in four important features. The first is the timestamp($t$) when the trade took place, with the precision in milliseconds. The second is the amount($a$) of asset that is transacted in the particular trade. For BTC-USDT trading pair, it is the amount of Bitcoin that changed hand in the trade. The third is the price($p$) in quote asset, in this case USDT, at which the transaction is settled. The last is a Boolean value, maker($m$), where "true" indicates the buyer of the trade put his/her order on the limit order book. Notice that since there is always a buyer and a seller when an asset changes hand, this Boolean value also indicates that whether this trade is the result of active selling or active buying. When this Boolean value is true, it means that a trader actively sold the asset to the buyer whose his/her order is on the limit order book, and vise versa. Specifically, each data point can be represented as a row vector: $d_{u}=[t(u), p(u), a(u), m(u)] \in \mathbb{R}^{1\times4}$ with $u$ representing the index of the data point in the data set.

\subsection{Trade-by-trade data representation in time intervals}
Since this study is interested in predicting the price movement in a fixed time horizon, it is critical to ensure there is a way to resample trade-by-trade data into fixed time intervals. Furthermore, using a representation in fixed time interval allow us to track a steady trailing window over time. Otherwise, if we choose to use 10,000 trades as input size, then it could mean trades happening in the last ten minutes during normal time or in the last thirty seconds in some extreme cases. This study adopts the following measures to resample trade-by-trade data into fixed time intervals with length $l$, measured in milliseconds. Start by group all the trades into subsets by the time interval they belong to. Specifically, one can obtain a group number $\textit{i}$ by the following operation: $i=\lfloor{\frac{
t(u)}{\textit{l}}}\rfloor$. The $k^{th}$ subset is essentially $S_{k} = \{d^{(i)}_u | i=k \}$. In each subset $S_{k}$, trade index is reset so that $u \in \{1,2,\dots,n\}$. After putting trades into subsets, the subsets are sorted in the order of group number.

Then, in each subset we have:
\begin{itemize}
\item $Number of Trades$ is simply defined as the count of total number of individual trades in the subset, $n$.
\item $\textit{Volume}$ is defined as the sum of amount in the subset.
    \begin{equation}
    \sum^{n}_{i=1} a_{i}
    \end{equation}
\item $\textit{Active Buy Volume}$ is defined as the sum of amount when maker equals to false.
    \begin{equation}
    \sum^{n}_{i=1} a_{i} \times (1-m_{i})
    \end{equation}
\item $\textit{Amplitude}$ is defined as the highest price minus the lowest price.
    \begin{equation}
    max\{p_i\} - min\{p_i\}
    \end{equation}
\item $\textit{Price Change}$ is defined as the price of the last trade minus the price of the first trade.
    \begin{equation}
    p_n - p_1
    \end{equation}
\item $\textit{Volume weighted average price}$ is defined as the sum of notional value of all trades divided by the total volume. The notional value of one trade is essentially the price of the trade times the amount of the trade.
    \begin{equation}
    {price} = \frac{\sum^{n}_{i=1} p_{i} \times a_{i}}{\sum^{n}_{i=1} a{i}}  
    \end{equation}
\item $\textit{Taker ratio}$ is defined as the sum of amount when maker equals to false divided by the total sum of amount.
    \begin{equation}
    \frac{\sum^{n}_{i=1} a_{i} \times m_{i}}{\sum^{n}_{i=1} a{i}}  
    \end{equation}
\end{itemize}

After restructuring individual trades into fixed time intervals, they will be sorted according to their group number $i$. Each data point $d^{'}_{i}$ in the new data structure now represents a fixed time interval with 10 features. The entire data set now is $D^{'} = [d^{'}_1, d^{'}_2, \dots, d^{'}_i, \dots,d^{'}_n]^T\in \mathbb{R}^{n\times7}$
This study focuses two values of \textit{l}, which are 60000 and 300000, corresponding to one minute intervals and five minute intervals respectively. For one minute time interval, the total number of intervals $n$ is 477,680, and for five minute intervals, it is 95,538. It is worthwhile to clarify how the features are found and selected. There are three principles the author follows in selecting these features: i) represent dimensions in original trade-by-trade data, ii) use relative measures to produce stationary features, iii) little correlation with other features. 

In addition to the input features, $\textit{change in next m interval}$ at time $t$, with $t$ represents a time-interval, an non-input feature which represents the price movement in a prediction horizon, is defined as the price at t+m divided by the price at t, and then minus one. Specifically,
    \begin{equation}
    C^{(m)}_{t}=\frac{{price}(t+m)}{{price}(t)} - 1
    \end{equation}
By varying parameter m, one can get the price movement in different prediction horizons. This feature will be computed before generating input data based on trailing windows so that it will be available for all data point later during labeling. This paper will focus on m=15,30 for one minute time interval, and 6, 24 for five-minute time interval.

\subsection{Stationary test}
Since time series data on asset prices are sensitive to regime changes, it is critical to check if the restructured data is stationary before divided them as input for the neural network. A piece of time series data is said to be non-stationary if it has unit-roots, indicating its values are time dependent. This study evaluates if the all features generated for the fixed time interval representations are stationary through the frequently-used Augmented Dickey–Fuller (ADF) test. The null hypothesis (H0) of the ADF test is that the time series has a unit root. Therefore, if the test results reject the null hypothesis, then the data is considered to be stationary. After running the ADF test on all the feature series, the author found all features are stationary,except for price, which has a test statistics of -1.454 and a p-value of 0.55. That test statistic is larger than the one for a 10\% confident interval, which is -2.566. Therefore, the price feature is adjusted by taking difference. Specifically, the new value for the feature is
\begin{equation}
    {price}^{'}(t)={price}(t)-{price}(t-1)
\end{equation}
After taking difference for price series, the new p-value is 0.0, and this show that it is now stationary.

\subsection{Normalization and labeling}
An input example can be generated in this ways: For prediction time $t>T$, a trailing window with length $T$ is the collection of data from $d_{t-T}$ to $d_{t}$. An input $X$ can now be defined as $X=[x_1,x_2,x_3,\ldots,x_t,\ldots,x_{T}]^T \in \mathbb{R}^{T\times4}$. Then a min-max normalization is applied to normalize all columns of input data. As defined earlier, the author will use $\textit{change in next m interval}$ to label different classes of price movement. Based on a threshold ($\epsilon$), the labels will be assigned as a row vector [1,0] for above the threshold, and [0,1] for below the threshold. The threshold will be slightly adjusted for different time horizon on which price movement to be predicted in order to achieve a well-balanced class distribution. It is worth emphasize the importance of having a balanced class distribution since it does not only ensure a fair baseline when interpreting the test result and comparing against random guesses, but also allow the model to emphasize each class equally in the process of learning their feature distributions. As we can see, the training data is well balanced and the baseline performance of a random guess should have about 50\% accuracy. Detailed class distribution is presented in TABEL II.

\begin{table}
\parbox{.45\linewidth}{
\centering
\begin{tabular}{|c|c|c|c|}
\hline
\textbf{m} & \textbf{$\epsilon$} & \textit{prev-dist.} & \textit{post-dist.}\\
\hline 
15 & 0.000 & 50.65\% & 50.65\%  \\
\hline 
30 & 0.000 & 50.80\% & 50.80\%  \\
\hline 
\end{tabular}
\vspace{-10pt}
}
\hfill
\parbox{.45\linewidth}{
\centering
\begin{tabular}{|c|c|c|c|}
\hline
\textbf{m} & \textbf{$\epsilon$} & \textit{prev-dist.}&\textit{post-dist.} \\
\hline 
6 & 0.0000 & 50.75\%  & 50.75\%\\
\hline 
24 & 0.0002 & 51.84\% & 50.07\% \\
\hline
\end{tabular}
\vspace{-10pt}
}
\vspace{10pt}
\caption{threshold value and class distribution}
\vspace{-10pt}
\end{table} 

\subsection{Training/testing allocation}
\subsubsection{Separation between Training and Validation}
The way to split the data into a training set and a validation set is described as follows: Randomly pick $p$ number of disconnected time intervals with length $q$ greater than trailing window length $T$ as validation periods. Then the entire data is left with $p+1$ number of disconnected time intervals, and those will be used as training periods. If a training period has length less than $T$, it will be discarded. This ensures that the input data for the model have a validation set and a training set that are generated from different time periods and have no intersection with one another. That is, the model has not ever peeked the data that is used for validation or testing when learning the parameters. If one failed to achieve this separation, then the model will have similar performance on the validation set and the training set due to the overlapping in the data, making the model likely to overfit the training set and not achieving satisfying performance on the out-of-sample data. This way of forming the training-validation separation has an important advantage in preserving the class distribution in the split data with a large $p$ and small $q$. The class distribution in training and validation periods are very close to the one of the whole training data. 
\subsubsection{Redundancy vs Quantity Trade-off}
After splitting the training and validation periods, input examples can be generated identically in each period. For a time periods with length q, the maximum number of input examples with intersection can be generated is $q-T$. These inputs have a great deal of overlapping. For example, a input that makes a prediction at time t will cover data from t to t-T, and an input at time t-1 will cover data from t-1 to t-1- T. Then T-1 number of vecotrs in these two inputs are essentially the same. To reduce such redundancy, one may wish to get inputs with no intersection with other inputs. Then the maximum number of inputs is $\lfloor{\frac{q}{\textit{T}}}\rfloor$. However, this leaves the model to little input examples to learn parameters. Therefore, it is crucial to balance the trade off between getting too many redundant training examples or having too little input examples. 
\subsubsection{Generate Input Examples by Offset Values}
The author approaches this problem by applying a number of offsets when getting non-intersect inputs in the same time period. When there is no offset, the first input is generated at t=T, so that the starting point of this input is at t = 0. Then the following inputs are generated at $t + n \times T $. Let us use an example to illustrate this. For a time period with length q=14, if one would want to get non-intersect input examples for trailing window T=4, then the eligible inputs are at t=4,8,12 with no offset. The maximum value of offset is T-1 = 3. If applying an offset value of 2, we will get inputs at t=6,10,14. In this way, one could expand the number of input examples very quickly with less overlapped data. 

In this study, the author randomly selects 10 to 50 percent from all eligible offset values and generated input examples in each time periods and the offset value of $period length \mod T$ is always selected to ensure a full coverage of the data. After iterate through all the training and validation periods, all input examples will be allocated to the training set or the validation set, respectively.
One could of course generate all eligible input data with sufficient computational power. Intuitively, having more data would not hurt if not improve the model's performance. However, sacrificing some input with great overlapping does not necessarily mean that the model could not learn a good set of parameters. The author choose to use only part of all the eligible offset value purely for the purpose of significantly reducing the work load during training while preserving a full coverage for the data.
\begin{figure}
    \centering
    \includegraphics[width=0.5\textwidth]{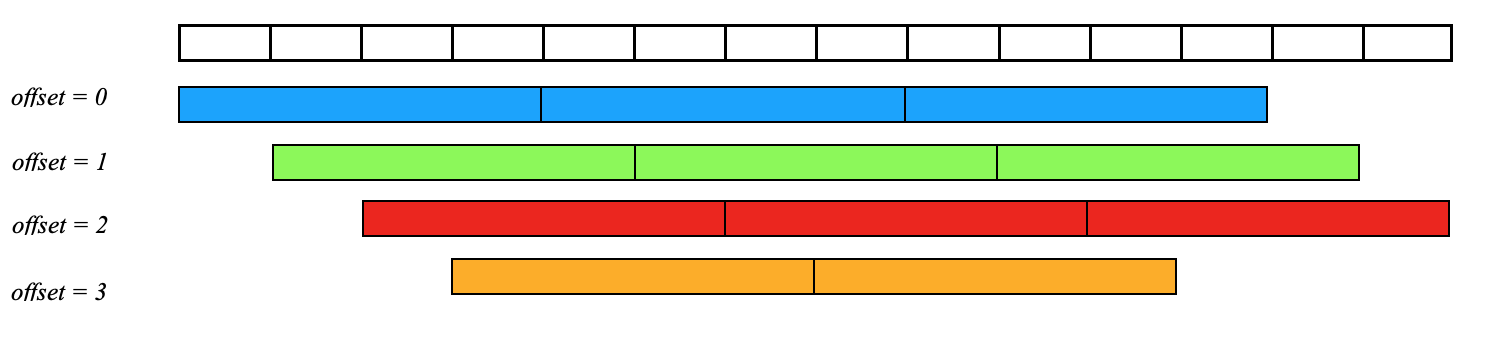}
    \caption{illustration of creating examples using offset}
    \label{fig:my_label}
\end{figure}

\section{Model Architecture}
 The model proposed in this study is a slightly altered version of a standard long short-term memory network(LSTM) network. The first part is an input layer, which takes in the normalized data. The next part is a layer of LSTM cells with length \textit{T}, connected with a drop-out layer with dropout rate of 0.5. The drop-out layer randomly selects half of output data from the LSTM layer and put them in the fully connected layer. The fully-connected layer with a softmax activation function outputs the prediction result at $t= T$.
 
\subsection{Long short-term memory network (LSTM)}
Recurrent neural networks has been proven useful in analyzing time series data and the LSTM network helps solve the vanishing gradient problem in training a recurrent neural network. Time series data are split into vectors $x_{t}$ and passed sequentially into LSTM cells at each time step. There are three types of gates in a LSTM cell--the update(input) gate, the output gate, and the forget gate. They are essentially governed by sets of trained parameters to maintain a cell state $c_{t}$, which could be passed through LSTM cells in parallel to the output of each cell $a_{t}$. Keeping a cell state allows the network to preserve information that is far away from the current time step and draw connections from long time dependencies. To generate the current cell state, we need the update gate and forget gate to decide whether each input information is used. It begins by stacking the output from previous cell $a_{t-1}$ and the raw data input of the current cell $x_{t}$ to form a combined input $A_{t}$. 
The update gate, governed by a set of parameter $W_{u}$ and bias term $b_{u}$, takes in the combined input $A_{t}$ and form a update gate values $\Gamma_{u}$, using the Sigmoid activation function. Specifically,
    \begin{equation}
    \Gamma_{u} = \sigma[W_{u} \times A_{t-1} + b_{u}]
    \end{equation}
The same procedure is done to generate the forget gate values $\Gamma_{f}$. The only difference is that it uses parameter $W_{f}$ and $b_{f}$. Specifically,
    \begin{equation}
    \Gamma_{f} = \sigma[W_{f} \times A_{t-1} + b_{f}]
    \end{equation}
Then by using a separated set of parameter $W_{c}$ and a bias parameter $b_{c}$ to evaluate the combined input and form $\tilde{c}_{t}$, using a \textit{tanh} activation function. 
    \begin{equation}
    \tilde{c}_{t} = tanh[W_{c} \times A_{t-1} + b_{c}]
    \end{equation}
With the update gate values dictating $tilde{c}_{t}$ and forget gate values dictating the previous cell state $c_{t-1}$, one can get the current cell state $c_{t}$. Specifically,
    \begin{equation}
    c_{t} = \Gamma_{u} \times \tilde{c}_{t} + \Gamma_{f} \times c_{t-1}
    \end{equation}
The output gate determines what information from the cell state is used when forming the output. One can get the output gate values $\Gamma_{o}$ using the same process for update gate values and forget gate values described above, with parameter $W_{o}$ and $b_{o}$. Specifically,
    \begin{equation} 
    \Gamma_{o} = \sigma[W_{o} \times A_{t-1} + b_{o}]
    \end{equation}
Then the output is simply obtained by evaluating the dot product of $\Gamma_{o}$ and the tanh-activated cell state $c_{t}$. Specifically,
    \begin{equation}
    a_{t} = \Gamma_{o} \times tanh[c_{t}]
    \end{equation}

\begin{figure}
    \centering
    \includegraphics[width=0.5\textwidth]{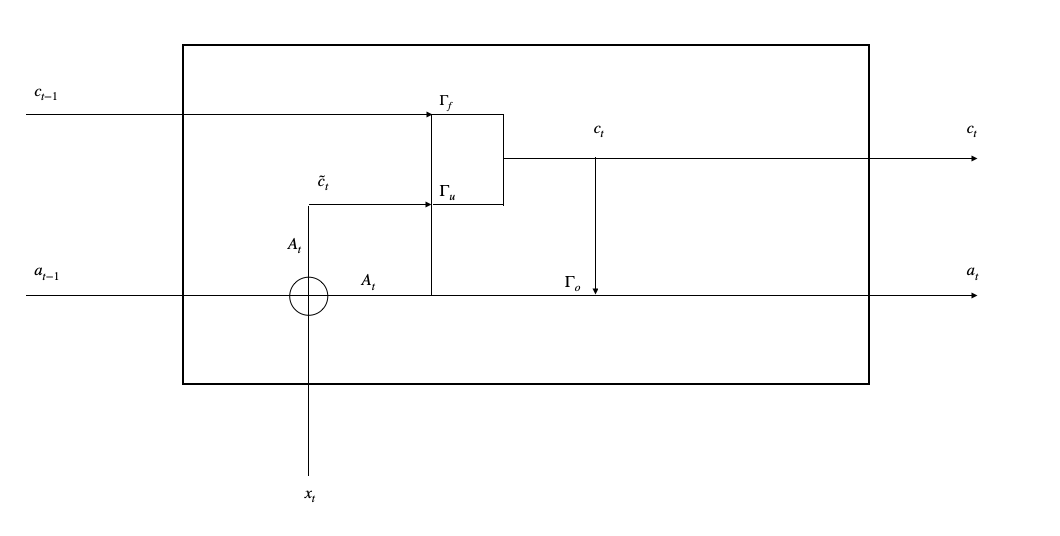}
    \caption{Structure of a LSTM cell}
    \label{fig:my_label}
\end{figure}
It is obvious that the length of the LSTM layer equals to size of the input data, which is the length of the trailing window $T$. Intuitively, the length of the trailing window determines how far the model will be able to look backward to predict the future price movement at the current time. It is a hyper-parameter that worth tuning because it can affect the training time of a model and the predictive capability. One would assume that with a longer trailing window, more time is required to train the model because the input contains more data and the model has more parameters. On the other hand, the model should make better prediction since more information is given to the model.

Another important hyper-parameter is the number of units(N) of the LSTM cell. This affects the dimensions of the LSTM cell output as well as the total parameter size of the model. Later, we will see that more parameter could more easily cause overfitting. Grid searches are performed in order to find the optimal values for these two hyper-parameters. For different prediction horizons and time interval length, the optimal choices of these two hyper-parameter are different. Detailed result and some interesting trends will discussed very soon.

\subsection{Training}
Before moving on to results of the grid search, it is worthwhile to describe the learning related setups and parameter choices. In this study, the model learns parameters by minimizing the categorical cross-entropy loss with the 'adam' optimizer that has initial learning rat of 0.001 and learning rate decays by 0.0003 for every 15 epochs until it reaches 0.0001. The mini-batch size ranges from 32 to 128, depending on the number of training and validation examples for different setups. A larger mini-batch will gradually decrease to a smaller size during the process of training. This choice is influenced by the finding in \cite{batch1}, that loss obtained from training with a small batch-size is more likely to converge to a shallow-board minimal. Another reason is that small batch-size tend to have a regularization effect, according to \cite{batch2}. Training is stopped if the validation loss does not improve for 20 or more epochs or if it has increased for more than 5$\%$ from the minimum value. In the latter case, the model is considered to be "over-fitting" the training set. Although lower values of loss usually translate into higher accuracy, this correlation does not hold for every epoch trained. In all the training process of this study, minimizing loss is prioritized over maximizing accuracy. The reason is that loss represents the general prediction capability of a model, which outputs close numerical values to assigned labels, while prediction accuracy is a result of such capability and it could be manipulated by setting different cut-off values in activation functions. Nevertheless, both metrics will be reported later. Moreover, the author chooses a softmax activation in the output layer to avoid tuning cut-off values. Training time ranges from a few hours for some models with small input size and fewer examples, and more than a day for large input size and more examples. All models are constructed by using Keras with the TensorFlow backend, and they are trained on a single NVIDIA GeForce GTX 1070 GPU. 

\subsection{Hyper-parameter Selection}
The grid search is performed separately for the two setups of time interval on two hyper-parameter T, and N. For one-minute time interval, the value range of T is 100,300,1000,2000, and the one of N is 16,32,64,128. For five-minute time interval, the value range of T is 60,300,500,1000, and the one of N is 16,32,64,128. For simplicity, the author will only report the best hyper-parameter pairs in terms of achieving the lowest validation loss for each setup in prediction horizon and time interval length in TABLE III. Fig.3 plots the learning curves for optimal hyper-parameters. Some interesting observations from the result are: i) The optimal number of neuron is 16 or 32. Greater number of units tends to over-fit the data earlier. ii)Prediction accuracy positively correlates with the length of the trailing window. However, the accuracy usually does not improve after an optimal value, suggesting that extra information in the past does not help make better prediction. iii) Smaller $m$ values and larger $l$ values tend to lead to better performance. By deduction, when predicting the same time horizon (notice that with l=60000,m=30 and l=300000,m=6, the model is predicting a thirty minute time horizon), larger $l$ values which corresponds to smaller m values leads to better performance. One potential reason could be that a larger l value compress information to a higher degree and noise is consolidated in this process and leaves clearer signals.

\begin{figure*}
    \centering
    \includegraphics[width=0.9\textwidth,height=8cm]{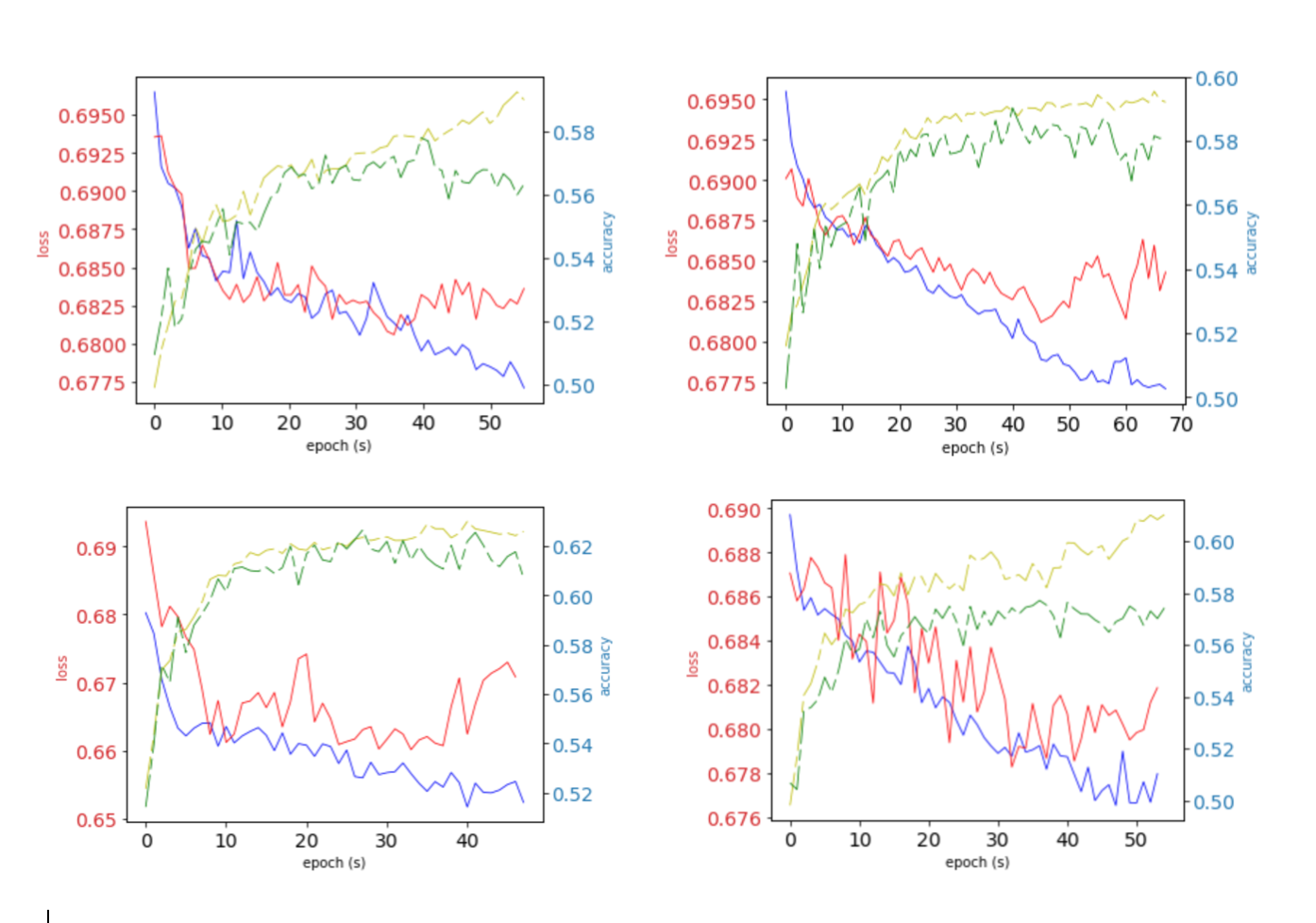}
    \caption{\textsf{ Blue: training loss, Red: validation loss, Yellow: training accuracy, Green: validation accuracy}}
    \label{fig:ds}
\end{figure*}

\begin{table}
    \centering
    \caption{Optimal hyper-parameter for setups}
    \vspace{5pt}
    \begin{tabular}{|c|c|c|c|c|}
    \hline
    \multicolumn{1}{|c|}{} &
    \multicolumn{2}{|c|}{l=60000}& \multicolumn{2}{|c|}{l=300000}\\ \hline
    m-value&m=15 & m=30 & m=6& m=24\\ \hline
    Na&16&32&16&16\\ \hline
    T&300&1000&300&500\\ \hline
    {Val\_loss}&0.6812 &0.6741 &0.6610 &0.6791 \\ \hline
    {Val\_acc}&{57.66\%} &{58.51\%} &{62.09\%} &{58.04\%} \\ \hline
    \end{tabular}
    \vspace{-10pt}
\end{table}

\section{Experimental Results}

This study conducts experiments on three out-of-sample data sets. These experiments can be breakdown into two categories, time interval length l=60000, and l=300000. For each time interval length setting, selected models described in the previous section will be used to make predictions on different setups in terms of future return horizons. Same as what was defined in Section 3, labels will be based on next m minutes return. For l=60000, those are m=15,30. And for l=300000, those are m=6,24.

\subsection{Experiment on BTC-USDT data set}
The BTC-USDT out-of-sample data set consist of all trades from 2019/12/01 to 2020/03/01. Same measures described in Section3 are utilized in aggregate these trades into fixed time intervals and in labeling them. In contrast to the practice in training our models, a threshold of 0 is chosen in producing all the labels. This indicates that there is no directional view on the out-of-sample and no effort in producing a balanced class distribution for the testing data. In fact, in a real trading environment, the model is very often dealing with a skewed distribution. Unlike the policy in training the model, which use a random subset of all valid input data, we will test our model on all the valid input data in a chronological order. By doing so, it not only helps to evaluate the accuracy comprehensively on all available test data, but also allows one to track a rolling accuracy, a metric which measures a model's performance on prediction accuracy in a rolling window of time. This metric gives insight into whether the performance is stationary in time and points out the regimes where the performance is particularly unsatisfying, allowing for further improvement. Ideally, one would wish the performance to be steady across the timeline. That will suggest the model's capability to constantly produce robust signal in real trading environment without any knowledge on the current regime. The overall test results are shown in TABLE IV and the distributions of daily rolling accuracy is shown in Fig. 4. In general, the high performance learned in training data translates well into the testing data. Setups with greater m value tends to loose more accuracy in the testing data. 
\begin{figure}
    \centering
    \includegraphics[width=0.4\textwidth]{boxplot.png}
    \caption{Rolling Accuracy Distribution}
    \label{fig:my_label}
\end{figure}

\begin{table}
    \centering
    \caption{Results on Out-of-Sample Test Periods}
    \vspace{5pt}
    \begin{tabular}{|c|c|c|c|c|}
    \hline
    \multicolumn{1}{|c|}{} &
    \multicolumn{2}{|c|}{l=60000}& \multicolumn{2}{|c|}{l=300000}\\ \hline
    m-value&m=15&m=30& m=6& m=24 \\ \hline
    loss&0.6845&0.6799&0.6720&0.6812\\ \hline
    accuracy&{57.18\%}&{57.65\%}&{61.12\%}&{57.08\%}\\ \hline
    \end{tabular}
    \vspace{-10pt}
\end{table}

\subsection{Experiment in a Trading Simulation}
Even though the focus of this paper is not to introduce a profitable trading strategy, rather it is to show the capability of the deep learning model in delivering robust price movement prediction, a trading simulation is presented below to show the predictions could easily be monetized. There are three assumptions. The first one assumes that trades in the trading simulation have no market impact, which means a trade made by the model does not affect trades from the data. One can achieve this by using a small order size relative to the existing orders on the limit order book. The second assumption is that there is no latency on execution. This implies that as soon as the model makes the prediction, it is able to execute the trade at the last price on that time interval. This condition is hardly achievable in real life, because there is friction and trade executions are always subject to the bid-ask spread at given time, although that spread is usually very small. Finally, a transaction cost of 0.0003 percent per order is applied to all trade executed. Such transaction cost is realistic in some instruments offered by exchanges.

The trading strategy is simply to take a long position when the model is predicting upward price movement and take a short position when the model is predicting the other direction. Note that the long positions and the short positions are separated. That is, taking a short position when already have a long position will not reduce the long position. The positions are closed only at the end of the prediction horizon, which means when predicting a five minute horizon, every position is held for five minutes. In reality, one can easily achieve so by using an exchange or broker that allows trading with isolated margin. For simplicity, the author runs only the best performing model on the test period in the trading simulation, which is resampling trade-by-trade data in five-minute time intervals with a look back periods of 300 and predicting the price movement in the next 5 intervals. The result is illustrated in Fig. 5.

\begin{figure}
    \centering
    \includegraphics[width=0.4\textwidth]{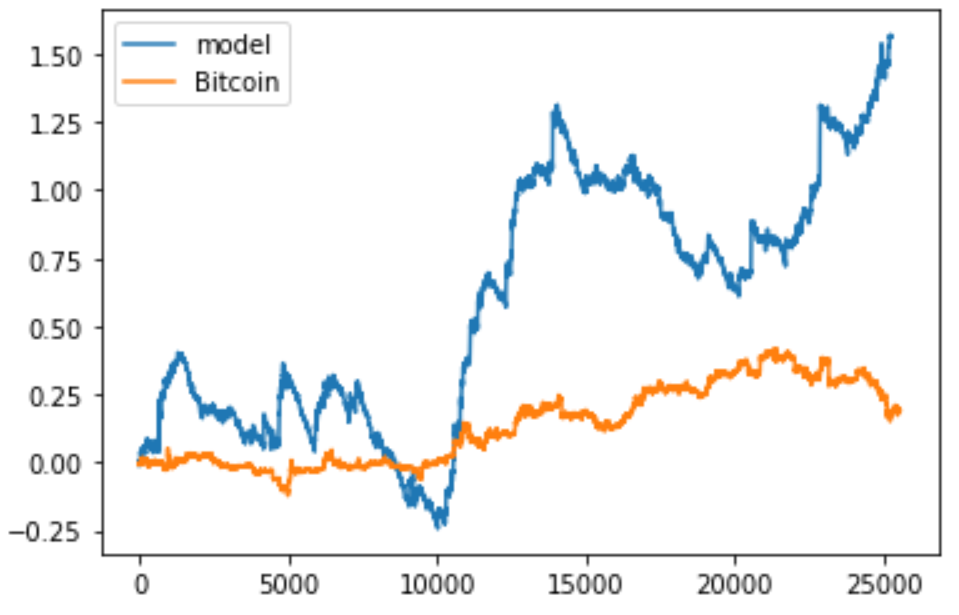}
    \caption{Trading simulation result}
    \label{fig:my_label}
    \vspace{-10pt}
\end{figure}

As we can see the trading strategy outperformed Bitcoin in terms of net return. However, it does have significant downturns in revenue. A possible reason is that the trading strategy is making trades in every interval and when the market is less active, the price movements in those periods are so small that it could not cover the transaction cost. Therefore, even if the trading strategy is making right prediction, it could potentially loose money. To reduce the number of trades and save transaction cost is indeed a direction of developing a more sophisticated strategy that utilize this prediction well.

\subsection{Experiment on other cryptocurrencies}
In this experiment, the model trained using BTC-USDT data is applied on other cryptocurrency trading pairs in order to examine that without additional training, if the learned parameters from one instrument can still be useful in predicting price movement of other instrument. In other word, this experiment could tell that if there exists universal features in trade-by-trade data from which the LSTM could extract to infer the direction of future prices. Among all digital assets, Ethereum(ETH), Bitcoin Cash(BCH), Litecoin(LTC) and EOS are top in terms of market capitalization and trading volume. Similar to the experiment on BTC-USDT, trade-by-trade data from the trading pairs of these cryptocurrencies against USDT are acquired in the time range from 2019/12/01 to 2020/02/28. The same experiment setup is adopted as the one on BTC-USDT data. For simplicity, only the best setup in BTC-USDT with l=300000 and m=6 is used in this experiment. The results are shown in Fig. 6. From this experiment, we can see that the prediction accuracy is very well translated to these trading pairs. Thus, it indicates that there is universal feature in trade-by-trade data across instruments from which the deep learning model could extract and utilize to make accurate predictions.

\begin{table}
    \centering
    \caption{Results on Out-of-Sample Test Periods}
    \vspace{5pt}
    \begin{tabular}{|c|c|c|c|c|}
    \hline
    Instrument & ETH & BCH & LTC & EOS \\ \hline
    accuracy&{60.48\%}&{60.17\%}&{59.96\%}&{60.03\%}\\ \hline
    \end{tabular}
    \vspace{-5pt}
\end{table}

\begin{figure}
    \centering
    \includegraphics[width=0.4\textwidth]{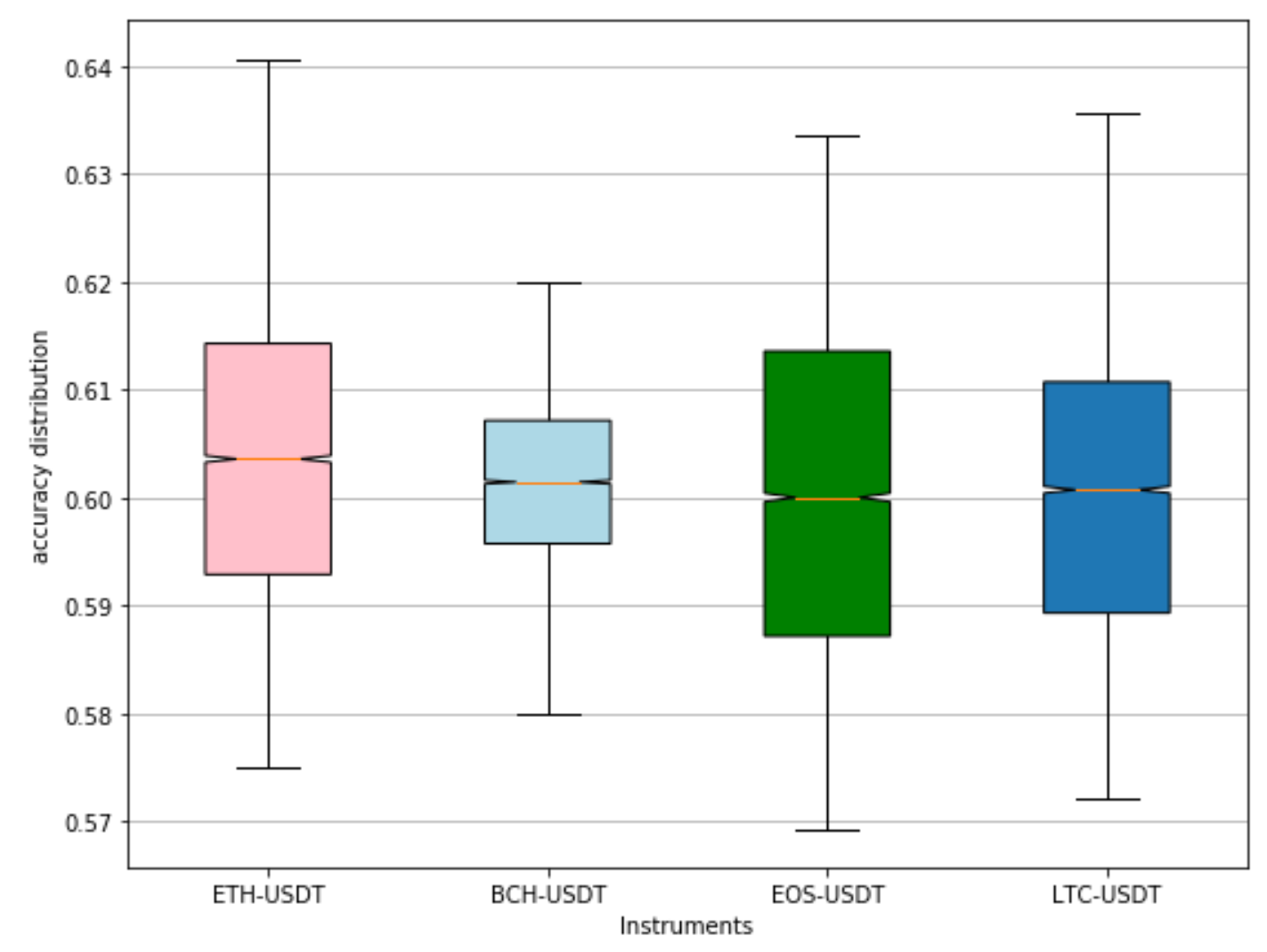}
    \caption{Accuracy Distribution on Other Digital Assets}
    \label{fig:my_label}
\end{figure}

\section{Conclusion and Future Works}

This study is the first to explore the application of a long short-term memory deep learning network on Bitcoin trade-by-trade data time series to make prediction on price movement in different horizons. With the most extensive and continuous price data, this research unveiled the relations between hyper-parameters and learning outcomes when applying an LSTM network to price data time series. This study also finds the correlation of model performance with the length of the resample time intervals, as well as the one with prediction horizon. Models with longer time interval and smaller prediction horizon tend to perform better. In addition, novel research practices is adopted in this article. First, the method for separating training data and validation data help preserve similar class distribution in the training set and the validation set. It also ensures the learned parameters are well transferred to the out-of-sample data. Second, the technique which involves using random offset values to generate input examples helps to significantly reduce redundancy and training workload while keeps full coverage and high utilization on data. Finally, using carefully designed feature, optimal hyper-parameter choices, and helpful training practices, the model produces satisfying results and successfully transferred the learned high prediction performance onto an out-of-sample test period over three months. Furthermore, the model maintains its good performance when tested against other instruments that is not part of the training data. This shows there exists universal feature which could be extracted by the deep learning framework presented in this study. Moreover, it suggests that the price of digital assets is to some degree predictable.

Although the models introduced in this studies are capable of delivering consistent high-accuracy prediction on Bitcoin price movement, they could provide no information on the magnitude of such movement. In fact, the magnitude is actually important in real trading. Therefore, the author will explore new models that are able to advise on the magnitude and direction of price movement. In addition, the author would also like to investigate in developing a sophisticated trading strategy that fully utilizes this model's prediction through means like optimizing the order size or determining the best holding time for each position, etc.

\bibliographystyle{project/IEEEtran}

\end{document}